\documentclass[aps,prd,preprint,floats,superscriptaddress]{revtex4}
\usepackage{graphicx}

\def\bwt{\begin{widetext}}
\def\ewt{\end{widetext}}
\def\be{\begin{equation}}
\def\ee{\end{equation}}
\def\bea{\begin{eqnarray}}
\def\eea{\end{eqnarray}}
\def\bean{\begin{eqnarray*}}
\def\eean{\end{eqnarray*}}
\def\bary{\begin{array}}
\def\eary{\end{array}}

\def\bit{\begin{itemize}}
\def\eit{\end{itemize}}

\def \ttb{\theta_{\tilde b}}

\begin{document}
\renewcommand{\thesection}{\Roman{section}}
\preprint{ANL-HEP-PR-02-055-Rev, EFI-02-90-Rev, hep-ph/0207235}
\preprint{November 2002}

\title{LIGHT GLUINO AND THE RUNNING OF $\alpha_s$}

\author{Cheng-Wei Chiang}
\email[e-mail: ]{chengwei@hep.uchicago.edu}
\affiliation{HEP Division, Argonne National Laboratory
9700 S. Cass Avenue, Argonne, IL 60439}
\affiliation{Enrico Fermi Institute and Department of Physics,
University of Chicago, 5640 S. Ellis Avenue, Chicago, IL 60637}
\author{Zumin Luo}
\email[e-mail: ]{zuminluo@midway.uchicago.edu}
\affiliation{Enrico Fermi Institute and Department of Physics, 
University of Chicago, 5640 S. Ellis Avenue, Chicago, IL 60637}
\author{Jonathan L. Rosner}
\email[e-mail: ]{rosner@hep.uchicago.edu}
\affiliation{Enrico Fermi Institute and Department of Physics, 
University of Chicago, 5640 S. Ellis Avenue, Chicago, IL 60637}

\date{\today}

\begin{abstract}
  A gluino in the mass range 12--16 GeV combined with a light (2--5.5 GeV)
  bottom squark, as has been proposed recently to explain an excess of $b$
  quark hadroproduction, would affect the momentum-scale dependence
  (``running'') of the strong coupling constant $\alpha_s$ in such a way as to
  raise its value at $M_Z$ by about $0.014 \pm 0.001$.  If one combines sources
  of uncertainty at low ($m_b$) and high ($M_Z$) mass scales, one cannot
  exclude such a possibility.  Prospects for improvement in this situation,
  which include better lattice QCD simulations and better measurements at
  $M_Z$, are discussed.

\end{abstract}


\maketitle

\section{INTRODUCTION \label{sec:int}}

The production of $b$ quarks in hadronic and electromagnetic reactions appears
enhanced with respect to expectations based on perturbative QCD
\cite{Frixione:2000si}.  While questions have been raised about the magnitude
or interpretation of this effect \cite{Cacciari:2002pa}, the discrepancy has
led to the suggestion of an additional mechanism of $b$ quark production
through the production of relatively light (12--16 GeV) gluinos, followed by
the decays of gluinos to $b$ quarks and their lighter (2--5.5 GeV)
superpartners $\tilde b$ \cite{Berger:2000mp}.  The orthogonal mixture $\tilde
b'$ is assumed to be sufficiently heavy that it would not yet have been
observed.  The $\tilde b$ squarks are assumed to be a mixture of the
superpartners of $b_L$ and $b_R$ such that the decay $Z \to \tilde b {\tilde
  b}^*$ is suppressed \cite{Carena:2000ka}.  Here we follow
\cite{Berger:2000mp,Berger:2001jb} in defining
\be
\left( \begin{array}{c} \tilde b \\ \tilde b' \end{array} \right) =
\left( \begin{array}{c c} \cos \ttb & \sin \ttb \cr
                        - \sin \ttb & \cos \ttb \end{array} \right)
\left( \begin{array}{c} \tilde b_R \\ \tilde b_L \end{array} \right)~~~.
\ee
We will see later that the light bottom squark $\tilde b$ is dominantly
right-handed in order to have a sufficiently weak coupling with the $Z$ boson.

A light gluino has been proposed before \cite{Fayet:1975pd,FF,Clav99}.
Clavelli \cite{Clav92} noted that the value of $\alpha_s$ as extracted from
quarkonia (see, e.g., Ref.\ \cite{KMRR}), when extrapolated using the standard
beta-function of QCD to $M_Z$, led to a slightly lower value than measured
directly at the $Z$.  The running effect can be slowed down through the
introduction of new fermionic and/or scalar particles with mass below $M_Z$.
Recent analyses do not exclude or favor the possibility of a light gluino in
the mass range of interest to us \cite{Bethke:2000ai,PDGQCD,Bethke:2002rv}.  (A
light gluino with mass of the order of a few GeV, however, has been
experimentally excluded \cite{deGouvea:1996va}.)  Ref.\ \cite{Becher:2001zb}
further shows that the inclusion of a light bottom squark only changes the
running slightly and is still compatible with the current experimental data.
However, the $\alpha_s$ extractions in these analyses do not take into account
the contributions of the new particles.  It is our purpose here to include such
effects using available results and identify the improvements in data and
calculations needed for a definite conclusion about the effect of a light
gluino on the running of $\alpha_s$ between scales below 10 GeV and the $Z$
mass.  This question is of interest because of foreseen improved determinations
at lower mass scales using quarkonium data and lattice gauge theories
\cite{Davies:1997mg,Davies:1998im,Davies:2002mv}, and at $M_Z$ using future
linear colliders.  We show that a distinction between the behavior of the QCD
beta function with and without a 12--16 GeV gluino is not possible at present,
but will be so with anticipated improvements in the low-energy determination of
$\alpha_s$ and with reduction on errors in $\Gamma(Z \to b \bar b)$ and
$\Gamma(Z \to \textrm{hadrons})$.

Section II treats two-loop formulae for the scale-dependence (``running'') of
$\alpha_s$ in the Standard Model (SM) and in the presence of a light gluino and
bottom squark.  Typical effects range from $\delta \alpha_s(M_Z) \equiv
\alpha_s^{MSSM}(M_Z) - \alpha_s^{SM}(M_Z) \simeq 0.015$ at $m_{\tilde g} = 12$
GeV to $\delta \alpha_s(M_Z) \simeq 0.009$ at $m_{\tilde g} = 30$ GeV, with
$\delta \alpha_s(M_Z) \simeq 0.002$ due to the bottom squark.  These effects
are somewhat larger than those found in Ref.\ \cite{Berger:2000mp} based upon
one-loop running, but errors on $\alpha_s$ at low mass scales (Section III), at
$M_Z$ (Section IV), and above $M_Z$ (Section V) still are large enough that no
distinction is possible between the Standard Model and the light-gluino/bottom
squark scenario in the minimal supersymmetric standard model (MSSM).  We
collect results and discuss the prospects for improved measurements in Section
VI, summarizing briefly in Section VII.

\section{TWO-LOOP RUNNING \label{sec:twoloop}}

The two-loop evolution of the strong coupling constant is governed by the
$\beta$ function
\begin{equation}
\beta(\alpha_s) = \mu \frac{d \alpha_s}{d \mu} =
  - \frac{\alpha_s^2}{2\pi}
    \left( b_1 + b_2 \frac{\alpha_s}{4\pi} \right) ~.
\end{equation}
In a minimally extended SUSY QCD model, the one- and two-loop coefficients
are given by \cite{Machacek:1983tz,Martin:1993yx,Clavelli:1996pz}
\begin{eqnarray}
\label{eqn:b1}
b_1 &=&
  \left( \frac{11}{3} - \frac23 n_{\tilde g} \right) C_A
  - \left( \frac43 n_q + \frac13 n_{\tilde q} \right) T_F ~, \\
b_2 &=&
  \left( \frac{34}{3} - \frac{16}{3} n_{\tilde g} \right) C_A^2
  - \left( 4n_q + 4n_{\tilde q} - 2n_{\tilde g}n_{\tilde q} \right) C_F T_F
  - \left( \frac{20}{3} n_q + \frac23 n_{\tilde q} - 2n_{\tilde g}n_{\tilde q}
    \right) C_A T_F ~,
\end{eqnarray}
where $n_q$ is the number of quark flavors, $n_{\tilde q}$ the number of
squarks, and $n_{\tilde g}$ the number of gluinos, $T_F = 1/2$ is the Dynkin
index of the fundamental representation, and $C_A = 3$ and $C_F = 4/3$ are the
Casimir invariants in the adjoint and fundamental representations,
respectively.  In the scheme with only one light bottom squark and one light
gluino with masses less than $M_Z$ ($n_{\tilde q} = n_{\tilde g} = 1$), the
changes in the $\beta$ function due to these new particles are
\begin{eqnarray}
&& \delta b_1^{\tilde g} = -2 ~, \qquad 
   \delta b_1^{\tilde b} = -\frac16 ~, \\
&& \delta b_2^{\tilde g} = -48 ~, \qquad 
   \delta b_2^{\tilde b} = -\frac{11}{3} ~, \qquad
   \delta b_2^{\tilde g \tilde q} = \frac{13}{3} ~.
\end{eqnarray}

Up to two loops, the decoupling relation between $\alpha_s^{(n_f)}(\mu)$ in the
$n_f$-flavor theory and $\alpha_s^{(n_f-1)}(\mu)$ in the $(n_f - 1)$-flavor
theory is trivial when they are matched at the heavy flavor threshold; for
example, $\alpha_s^{(n_f)}(m_b) = \alpha_s^{(n_f-1)}(m_b)$ for the
$\overline{\rm MS}$ mass $m_b = m_b^{(n_f)}(m_b)$.  Finite corrections start to
come in when one considers three-loop running \cite{Chetyrkin:1997un}.

Starting from $\alpha_s$ at a low energy scale, one can obtain its value at
$M_Z$ by solving the integral equation
\begin{equation}
\log\left( \frac{M_Z^2}{\mu_0^2} \right) =
  \int_{\alpha_s(\mu_0)}^{\alpha_s(M_Z)} \frac{2 d \alpha}{\beta (\alpha)} ~.
\end{equation}
Evolving the strong coupling constant in the SM and MSSM with initial values
given in Ref.~\cite{Davies:2002mv} at $m_b=4.1$ GeV, $\alpha_s^{(n_f=5)}(m_b) =
0.239 ^{+0.012}_{-0.010}$, we take $m_{\tilde b} = 4$ GeV and $m_{\tilde
g} = 15$ GeV as an example and obtain
\begin{equation}
  \alpha_s^{\rm SM} (M_Z) = 0.1216 \pm 0.0027 ~, \qquad
  \alpha_s^{\rm MSSM} (M_Z) = 0.1352 \pm 0.0035 ~.
\label{alphasMSSM}
\end{equation}
It should be mentioned that the minor difference of the evolution within the SM 
of this paper from that given in Ref.~\cite{Davies:2002mv} is because we
restrict ourselves to two-loop running while they use the three-loop running
result.

\begin{figure}
\includegraphics{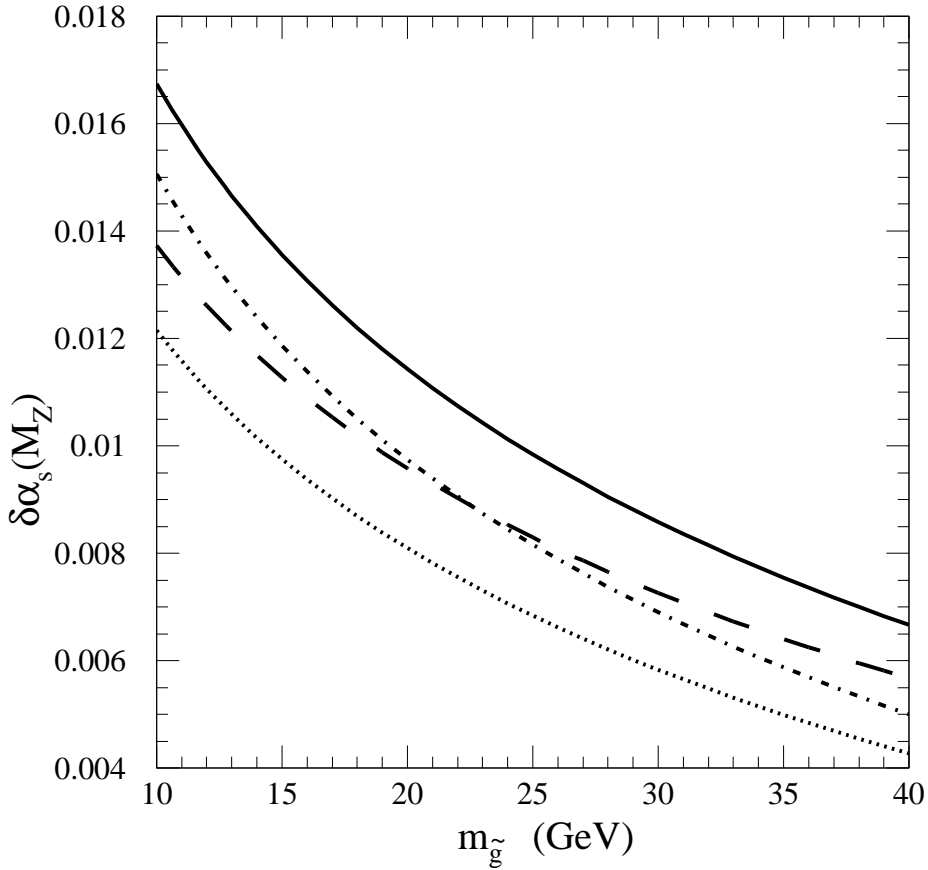}
\caption[]{\label{fig:dtas}
  Difference of $\alpha_s (M_Z)$ between the MSSM and SM starting from
  $\alpha_s (m_b) = 0.239$ as a function of the gluino mass $m_{\tilde g}$.
  The solid and dot-dashed curves give the two-loop results with and without a
  light bottom squark, respectively.  The corresponding one-loop results are
  shown by the long-dashed curve (gluino plus light bottom squark) and dotted
  curve (gluino only).}
\end{figure}

We find that the light gluino dominates over the light bottom squark in the
evolution over a wide range of its mass.  In Fig.~\ref{fig:dtas}, we plot the
difference $\delta \alpha_s (M_Z)$ as a function of the gluino mass $m_{\tilde
 g}$.  The solid curve gives the result with both the light bottom squark and
gluino taken into account, whereas the dashed curve gives the result due to the
light gluino contribution alone.  The corresponding one-loop running results
are indicated by the long-dashed curve (gluino and bottom squark) and dotted
curve (gluino only).
For the range of gluino mass of interest, 12 to 16 GeV, we find $\delta
\alpha_s$ ranges from 0.015 to 0.013, so we shall quote it as $\delta \alpha_s
= 0.014 \pm 0.001$ in what follows.  We now ask whether present data favor or
disfavor such an effect.

\section{LOW-ENERGY INFORMATION ON $\alpha_s$}

In this section we review the main sources of low-energy information on
$\alpha_s$, concentrating on those with the smallest claimed errors and
describing those errors critically.  All results are quoted as $\alpha_s(M_Z)$
assuming SM running in this section. The corresponding values of
$\alpha_s(M_Z)$ in the presence of a light gluino and a bottom squark can be
obtained by adding $\delta \alpha_s~(=0.014 \pm 0.001)$.

\subsection{$\tau$ decays}

The lowest-energy determination of $\alpha_s$ which appears in current reviews
\cite{Bethke:2000ai,PDGQCD,Bethke:2002rv} comes from $\tau$ decays involving
hadrons, with a {\it maximum} mass scale of $m_\tau$.  Impressive progress has
been claimed in expressing the hadronic final state in $\tau \to \nu_\tau + X$,
$X = \pi, \rho, a_1, \ldots$, in terms of an effective quark-antiquark
continuum describable via perturbative QCD, leading to values $\alpha_s(m_\tau)
= 0.323 \pm 0.030$ \cite{Bethke:2000ai,Bethke:2002rv} or $0.35 \pm 0.03$
\cite{PDGQCD}.  Extrapolation via the renormalization group then leads to the
values $\alpha_s(M_Z) = 0.1181 \pm 0.0031$ \cite{Bethke:2000ai,Bethke:2002rv}
or $0.121 \pm 0.003$ \cite{PDGQCD}.  However, the assignment of errors to the
contribution of nonperturbative effects in these analyses is highly subjective,
based on QCD sum rules for which independent tests of sufficient accuracy do
not exist in our opinion.  Remember that an error of 9\% in $\alpha_s(m_\tau)$
corresponds to a change of less than 3\% in the perturbative expression for
$\Gamma(\tau \to \nu_\tau + {\rm hadrons})$.

\subsection{Deep inelastic scattering}

The original and still one of the most powerful methods to measure $\alpha_s$
is deep inelastic lepton-hadron scattering (DIS).  The most precise of several
determinations, based on measurements of the structure function $F_2$ with
electrons and muons, gives $\alpha_s (M_Z) = 0.1166 \pm 0.0022$
\cite{Bethke:2002rv}, from data points in the range $1.9 \, {\rm GeV} \le Q \le
15.2 \, {\rm GeV}$ \cite{Santiago:2001mh}.  However, the error is not based on
an analysis by any experimental group.  Other determinations are consistent
with this value, but the smallest error quoted in any of them is $\pm 0.004$
\cite{Bethke:2000ai,Bethke:2002rv}.

In principle the determination of $\alpha_s$ from deep inelastic scattering
could be affected by a light $\tilde b$, since at the highest $Q^2 \gg
m_{\tilde b}^2$ gluons can split into $\tilde b \tilde b^*$, affecting the
evolution equations.  However, such an analysis is beyond the scope of the
present note, and is more appropriately carried out by the experimental groups
themselves.
 
\subsection{Quarkonium}

The measurement of $\alpha_s (m_b) = 0.22 \pm 0.02$ from the $\Upsilon$ system
(for $m_b = 4.75$ GeV) implies $\alpha_s (M_Z) = 0.118 \pm 0.006$ in Bethke's
review \cite{Bethke:2000ai,Bethke:2002rv}.  (A lower value $\alpha_s(m_b)=0.185
\pm 0.01$, implying $\alpha_s(M_Z)=0.109 \pm0.004$, is quoted by Hinchliffe in
the Particle Data Group review \cite{PDGQCD}.) Neither value is competitive in
its errors with the most precise one based on deep inelastic lepton-hadron
scattering.

The presence of light $\tilde b$ squarks affects the determination of
$\alpha_s$ from certain quarkonium decays.  For example, the total width of the
$\Upsilon$ is affected if the decay $\Upsilon \to \tilde{b} {\tilde b}^*$ is
permitted.  One has \cite{Berger:2001jb}
\begin{equation}
R^{\Upsilon}_{\tilde{b} {\tilde b}^*} =
  \frac{\Gamma_{\tilde{b} {\tilde b}^*}}{\Gamma_{\ell \bar{\ell}}} =
  \frac{1}{3} \left(\frac{\alpha_s(\mu)}{\alpha_{\rm{em}}}\right)^2
    \frac{m_{\Upsilon} (m_{\Upsilon}^2-4m_{\tilde{b}}^2)^{3/2}}
         {(t-m^2_{\tilde{g}})^2} ~,
\label{ratio}
\end{equation}
where $t = -(m_{\Upsilon}^2 - 4 m_{\tilde g}^2) / 4$.  Typical effects on the
bottom squark partial widths can exceed ten times the leptonic widths, thereby
attaining values of tens of keV for different $\Upsilon$ states and
substantially affecting their expected total widths.  Suppose the bottom
squarks in the final state behave like usual hadronic jets within the detector.
To compensate for the new open channel of $\tilde b {\tilde b}^*$, for a given
{\it measured} hadronic width one must reduce the value of $\alpha_s (m_b)$.
From Eq.~(\ref{ratio}) and $\Gamma_{\rm had}^{\Upsilon(1S)} = 52.5 \pm 1.8$
keV, we find that $R^{\Upsilon}_{\tilde{b} {\tilde b}^*} \simeq 9$ for
$m_{\tilde b} = 4$ GeV, $m_{\tilde g} = 14$ GeV and $\alpha_s$ should be
reduced by about $5\%$, consistent with the estimate in Ref.\ 
\cite{Berger:2001jb}.  Such a change is well within the current error on the
extracted $\alpha_s (m_b)$.

The decays of $\chi_{bJ}$ states to $\tilde{b} {\tilde b}^*$ occur with partial
widths which can exceed 200 keV for $J=0$ and for $\sin \theta_{\tilde b} \cos
\theta_{\tilde b} > 0$; for $J=1,~2$ these partial widths are calculated to be
much smaller, and for the other sign of $\sin \theta_{\tilde b} \cos
\theta_{\tilde b}$ they are small for all three values of $J$.  One must then
take account of these changes when extracting $\alpha_s$ from $\chi_{bJ}$ data,
but their impact on the determination mentioned above is relatively modest
\cite{Berger:2002gu}.

\subsection{Lattice}

We shall argue that lattice calculations of $\alpha_s$ using the upsilon levels
are relatively insensitive to the new physics introduced by light bottom
squarks and gluinos.  The inputs to the calculation of Ref.\ 
\cite{Davies:1997mg} are an overall mass scale (essentially adjustable through
the choice of $m_b$) and either a $2S$--$1S$ or a $1P$--$1S$ level spacing.
Both level spacings are used to obtain a value of $\alpha_s$ at a low mass
scale characteristic of $m_b$.  Consistency between the two values is used to
argue in favor of an unquenched calculation with $n_f = 3$ light quark flavors.

The new open $\tilde b {\tilde b}^*$ decay channels affect not only the decay
widths, but also the masses of the $b \bar b$ bound states.  In analogy with
the neutral kaon system, in which the $K_L$--$K_S$ mass splitting is of the
same order as the $K_S$ decay width to $\pi \pi$, we shall assume that the mass
shifts in $b \bar b$ bound states due to the open $\tilde b {\tilde b}^*$
channels are of the {\it same order} as the contributions of $\tilde b {\tilde
  b}^*$ decay channels to their partial widths, i.e., tens of keV for the
S-wave levels, at most a couple of hundred keV for the $^3P_0$ level, and
unimportant for the other P-wave levels.  A potential contribution from the
heavy bottom squark ${\tilde b}^{\prime}$ is estimated to be unimportant
because of the large mass suppression.  The spin-weighted average
$[5M(\chi_{b2}) + 3M(\chi_{b1}) + M(\chi_{b0})]/9 = \bar M(1P)$ is then
affected by at most tens of keV.  This is to be compared with the input level
spacings $M(\Upsilon') - M(\Upsilon) = 563$ MeV and $M(\chi_b) - M(\Upsilon) =
440$ MeV \cite{Davies:1997mg}.  We evaluate the effect of shifts in these
quantities by tens of keV on $\alpha_s$ as follows.

A scale change $r$ in the input mass splittings is reflected in a similar
change in the scale at which $\alpha_s$ is evaluated, $\alpha_s(M) \to
\alpha_s(Mr)$.  Using this estimate, we find that the change in $\alpha_s$ due
to a change by a factor of $r = 1 + \delta$ is $\Delta [\alpha_s^{-1}] \simeq
(b_1/2 \pi) \delta$, where $b_1(n_f=5, \tilde b) = 7.5$ from Eq.~(\ref{eqn:b1})
so $\Delta [\alpha_s^{-1}] \simeq \delta \simeq 10^{-4}$, $\Delta \alpha_s
\simeq 10^{-4} \alpha_s^2$.  This is smaller by orders of magnitude than the
effects which we consider to be important.

The lattice calculation of $\alpha_s$ at scales of order $m_b$ is thus not
likely to be affected by the presence of a light gluino and bottom squark with
the mass ranges considered here to more than $\pm 0.0001$, and possibly even
greater accuracy.

The Particle Data Group review by Hinchliffe \cite{PDGQCD} quotes the average
of several lattice determinations as implying $\alpha_s(M_Z) = 0.1134\pm
0.003$, at a characteristic mass scale of $m_b$ as in the case of quarkonium
decays.  Bethke \cite{Bethke:2002rv} adopts only the latest lattice
determination \cite{Davies:2002mv} and quotes $\alpha_s(M_Z) = 0.121 \pm
0.003$.

\section{INFORMATION ON $\alpha_s$ AT $M_Z$}

\subsection{Direct measurements: Standard Model}

Based on $\alpha_s(M_Z)=0.1200 \pm 0.0028$ and the global best fit values of
some other input parameters (e.g., $M_Z$, $M_H$, etc.), the Standard Model
predicts $\Gamma(Z \to {\rm hadrons})=1.7429 \pm 0.0015$ GeV, to be compared
with the experimental value $1.7444 \pm 0.0020$ GeV \cite{PDGEL}.  (The fact
that the two numbers do not agree exactly is due to the existence of other
inputs in the fit affected by $\alpha_s(M_Z)$.)  The experimental error alone
in $\Gamma(Z \to {\rm hadrons})$ would imply an error in $\alpha_s(M_Z)$ of
$\pm 0.0034$, consistent with the value quoted by Bethke \cite {Bethke:2002rv}.
Additional theoretical errors raise this to $\pm 0.005$.  Fitting {\it only}
$\Gamma(Z \to {\rm hadrons})$, we find $\alpha_s(M_Z)=0.123 \pm 0.005$.  We
shall adopt this more conservative error.

\subsection{Effect of SUSY scenario on $\Gamma(Z \to \tilde b {\tilde b}^*)$}

The light bottom squark $\tilde b$ is assumed to be long-lived at the collider
scale or to decay promptly into light hadrons in this scenario
\cite{Berger:2000mp}.  In either case, it forms a hadronic jet within the
detector due to its color charge.  Therefore, the $Z \to \tilde b {\tilde b}^*$
decay mode will contribute to the total hadronic width of the $Z$ boson.

The partial decay width $\Gamma(Z \to \tilde b {\tilde b}^*)$ can be expressed
at the tree level as
\be
\Gamma(Z \to \tilde b {\tilde b}^*) \simeq \frac{G_F
M_Z^3}{8\sqrt{2}\pi}\left[(g_V^b+g_A^b)\sin^2\theta_{\tilde
b}+(g_V^b-g_A^b)\cos^2\theta_{\tilde b}\right]^2 , 
\label{eqn:Zsbsb}
\ee
where we take the limit $m_{\tilde b} \approx 0$.  (In our convention the $Z b
{\bar b}$ vertex $\sim$ $g_V^b - g_A^b \gamma^5$.)  The $Z {\tilde b} {\tilde
  b}^*$ coupling must be small to agree with the electroweak precision
measurements at the $Z$-pole. A vanishing tree-level $Z \tilde b {\tilde b}^*$
coupling is achieved if the mixing angle $\theta_{\tilde b}$ is chosen to
satisfy $\sin \theta_{\tilde b}=\sqrt{2\sin^2\theta_W/3} \simeq 0.39$. However,
a nonzero effective coupling, which can be obtained if $\sin \theta_{\tilde b}
\ne 0.39$ and/or via loop corrections, may contribute to $\Gamma(Z \to \tilde b
{\tilde b}^*)$. M. Carena {\it et al} \cite{Carena:2000ka} calculated the
$\tilde b {\tilde b}^*$ production cross section as a function of the effective
$Z \tilde b {\tilde b}^*$ coupling.  Their results indicate that $\Gamma(Z \to
\tilde b {\tilde b}^*)$ is less than ${\cal O}(0.001)$ GeV for $0.30 \leq \sin
\theta_{\tilde b} \leq 0.45$. For comparison, the tree-level formula
(\ref{eqn:Zsbsb}) gives $\Gamma(Z \to \tilde b {\tilde b}^*)=(0 \sim 0.001)$
GeV in the same range of $\sin \theta_{\tilde b}$. For $\sin \theta_{\tilde b}
\simeq 0.39$, an upper bound can be obtained on the one-loop correction to
$\Gamma(Z \to \tilde b {\tilde b}^*)$ by using an argument similar to that in
Section\ \ref{sec:Zsgsg}, which would also assert that $\Gamma(Z \to {\tilde b}
{\tilde b}^*)$ is less than ${\cal O}(0.001)$ GeV.

\subsection{Effect of SUSY scenario on $\Gamma(Z \to b \bar b)$
\label{sec:Zbb}}

The electroweak observables such as $R_b$ have been considered to provide a
stringent constraint on the allowed parameter space of the light gluino/bottom
squark scenario \cite{Cao:2001rz,Cho:2002mt,Baek:2002xf}.  For $\sin
\theta_{\tilde b}=0.39$, $m_{\tilde b} = 5$ GeV, $m_{\tilde b'}=200$ GeV, and
$m_{\tilde g}=14$ GeV, S. Baek \cite{Baek:2002xf} calculated $\delta R_b \equiv
R_b - R_b^{\rm SM}$ as a function of the CP violating phases $\phi_b$ and
$\phi_3$. The range of $\delta R_b$ turns out to be $-$(2.0 -- 3.5) $\times
10^{-3}$ for $\sin\theta_{\tilde b}=0.39$ and $m_{{\tilde b}^*}=200$ GeV. This
is unacceptably large.  The observed value is $R_b^{\rm expt} = 0.21664 \pm
0.00068$ \cite{PDGEL}, to be compared with the Standard Model prediction
$R_b^{\rm SM} = 0.21569 \pm 0.00016$.  The difference is $R_b^{\rm expt} -
R_b^{\rm SM} = 0.00095 \pm 0.00070$, so that one must have $3.05 \times 10^{-3}
> \delta R_b > -1.15 \times 10^{-3}$ to maintain agreement at the $3 \sigma$
level.  By suitable choice of phases Baek is able to reduce the predicted
magnitude of $\delta R_b$ by about a factor of two, which would put it within
reasonable limits.  Dealing with the CP conserving MSSM, J. Cao {\it et al.}
\cite{Cao:2001rz} obtained similar results for $m_{\tilde b}=3.5$ GeV.

It is noted in Refs.\ \cite{Baek:2002xf, Cao:2001rz} that the variation of
$m_{\tilde b}$ does not change $\delta R_b$ significantly.  As will be seen
later, the SUSY contribution to the decay channel $Z \to b \bar b$ is the
dominant component in the change of the hadronic $Z$ decay width.  Therefore,
we take $\delta \Gamma(Z \to {\rm hadrons}) \simeq \delta \Gamma(Z \to b {\bar
  b})$, which is related to $\delta R_b$ by
\bea
\delta R_b 
&\simeq&
  \frac{\delta \Gamma(Z \to b {\bar b})}
       {\Gamma^{\rm SM}(Z \to {\rm hadrons})}
- \frac{\Gamma^{\rm SM}(Z \to b {\bar b}) \delta \Gamma(Z \to {\rm hadrons})}
       {\Gamma^{\rm SM}(Z \to {\rm hadrons})^2} \nonumber \\
&=& (1-R_b^{\rm SM})\frac{\delta \Gamma(Z \to b {\bar b})}
                         {\Gamma^{\rm SM}(Z \to {\rm hadrons})} ~.
\eea

In the following calculation, we will take the range $\delta R_b = (-1 \sim -2)
\times 10^{-3}$ (which covers the most negative acceptable value if one takes
the current $3 \sigma$ bound seriously) for our estimation of changes in
$\alpha_s(M_Z)$.  Using $R_b^{\rm SM}=0.21569$ and $\Gamma^{\rm SM}(Z \to {\rm
  hadrons})=1.7429$ GeV, one finds that $\delta \Gamma(Z \to b {\bar
  b})=-(0.0022 \sim 0.0044)$ GeV.  Here we reiterate that the value of $R_b$
predicted in the SUSY scenario remains a potentially dangerous feature of this
scheme.
 
\subsection{Effect of SUSY scenario on $\Gamma(Z \to {\tilde g}{\tilde
g})$ \label{sec:Zsgsg}}

With a light gluino in this scenario, the $Z$ boson can decay into a pair of
gluinos through loop-mediated processes.  The gluinos then decay promptly to $b
{\tilde b}^*$ or ${\bar b}{\tilde b}$, contributing to the total hadronic width
of the $Z$.  Previous analyses \cite{Kane:xp} indicate that the branching ratio
of $Z \to {\tilde g}{\tilde g}$ falls in the range of $10^{-5}$ to $10^{-4}$
for a wide range of MSSM parameter space.  This gives a partial width of less
than ${\cal O}(1)$ MeV.  Although the possibility of a light bottom squark is
not considered in those analyses, it can be argued that any possible increase
due to the light bottom squark should be comparatively small.  The reason is
that the effective coupling between the $Z$ boson and the gluinos should be of
the same order as the one-loop correction to the coupling between $Z$ and
bottom quarks, both of which are $\alpha_s$ and one-loop suppressed.  The
$Z$$b$${\bar b}$ coupling receives an ${\cal O}(\alpha_s)$ correction coming
from the interference between the SUSY contribution and the SM tree-level
coupling, resulting in a decrease of at most $4.4$ MeV in the total width of
$Z$ (see Section \ref{sec:Zbb}).  In the case of $Z$${\tilde g}$${\tilde g}$,
however, there is no tree-level coupling; therefore, the amplitude for the
process is further suppressed by ${\cal O}(\alpha_s)$.  Using the result of
$\delta\Gamma(Z \to b \bar b)$ as given in Section \ref{sec:Zbb}, it is easy to
see that the partial width of $Z \to {\tilde g}{\tilde g}$ is indeed at most an
MeV.

A lower bound can be obtained on $\Gamma(Z \to {\tilde g}{\tilde g})$ based on
the unitarity of the $S$-matrix ($S^{\dagger} S = 1$).  We expect that this
bound is likely to provide a fairly good estimate of the actual $Z \to {\tilde
  g}{\tilde g}$ partial width as long as cancellations of loop contributions
with high internal momenta are implemented, as in the calculations of Ref.\ 
\cite{Kane:xp}.  The situation is analogous to the $K_S$--$K_L$ mass difference
and the decay $K_L \to \mu^+ \mu^-$.  In each case the high-momentum components
of the loop diagrams are suppressed (here, through the presence of the charmed
quark \cite{GL}), leaving the low-mass on-shell states ($\pi \pi$ or $\gamma
\gamma$, respectively) to provide a good estimate of the matrix element.

The imaginary part of the invariant matrix element ${\cal M}(Z \to {\tilde
  g}{\tilde g})$ can be written as
\be
{\rm Im} \left[{\cal M}(Z \to {\tilde g}{\tilde g}) \right] =
\frac{1}{2} \sum_f \int d\Pi_f {\cal M}(Z \to f){\cal M}^*({\tilde
g}{\tilde g} \to f) ,
\ee
where the sum runs over all possible intermediate on-shell states $f$. Since
${\tilde b}$ is the lightest supersymmetric particle in the scenario and all
other supersymmetric particles (except ${\tilde g}$) are much heavier, we only
need to consider the cases where $f$ is $b {\bar b}$ and ${\tilde b} {\tilde
  b}^*$.  The contribution of the latter can be neglected because we require
that the tree-level $Z{\tilde b}{\tilde b}^*$ coupling be small.  Furthermore,
since the mass of the heavy sbottom ${\tilde b}^{\prime}$ is very large, only
${\tilde g}{\tilde g} \to b{\bar b}$ via ${\tilde b}$ exchange is considered to
be significant.  Based on the fact that $|{\cal M}(Z \to {\tilde g}{\tilde g})|
\ge {\rm Im} \left[ {\cal M}(Z \to {\tilde g}{\tilde g}) \right]$, our
calculation indicates \cite{wip}
\be
\frac{\Gamma(Z \to {\tilde g}{\tilde g})}{\Gamma(Z \to b {\bar b})} \ge
\frac{\alpha_s^2(M_Z)}{24}\frac{(g_V^b+g_A^b)^2 \sin^4\theta_{\tilde
b}+(g_V^b-g_A^b)^2 \cos^4\theta_{\tilde b}}{(g_V^b)^2+(g_A^b)^2} ,
\ee
Taking $\alpha_s(M_Z)=0.123$ and $\sin\theta_{\tilde b}=0.39$, we obtain
$\Gamma(Z \to {\tilde g}{\tilde g}) \ge 0.02$ MeV. As mentioned above, it is
likely that the actual partial width is not far above this lower bound.  We
will take the upper bound to be 1 MeV, as explained earlier.

\vspace{2ex}
In summary, we estimate $\Gamma(Z \to {\tilde b}{\tilde b}^*)=(0 \sim 1)$ MeV,
$\delta\Gamma(Z \to b {\bar b}) = -(2.2 \sim 4.4)$ MeV and $\Gamma(Z \to
{\tilde g}{\tilde g}) = (0.02 \sim 1)$ MeV. The total correction to the
predicted hadronic width of $Z$ is thus $(-4.4 \sim -0.2)$ MeV, which is
equivalent to a change of $(0 \sim +0.008)$ in $\alpha_s(M_Z)$ with respect to
the SM value $0.123 \pm 0.005$.  We then have $\alpha_s(M_Z) = (0.123 \sim
0.131) \pm 0.005$ in the SUSY scenario.

\section{INFORMATION ON $\alpha_s$ ABOVE THE $Z$}

A number of determinations of $\alpha_s$ at the highest-available mass scales
are based on event shapes in $e^+ e^-$ annihilations \cite{Bethke:2002rv}.  An
example \cite{Achard:2002jb} of such determinations, based on data at
center-of-mass energies up to 206 GeV, is $\alpha_s(M_Z) = 0.1227 \pm 0.0012
\pm 0.0058$.  Since the dominant error is systematic, it will not be decreased
substantially by combination with results of other experiments.

The determination of $\alpha_s(M_Z)$ from event shapes in high-energy $e^+ e^-$
annihilations will be affected in several ways by the light-gluino/light bottom
squark scenario.  Virtual bottom quarks will be able to radiate bottom squarks
and gluinos; virtual gluons will be able to split into pairs of light bottom
squarks and pairs of gluinos; and NNLO perturbative expressions will be
modified because of new loops in gluon and bottom quark propagators.  Estimates
of these effects are beyond the scope of the present note, but are worth
pursuing.

\section{RESULTS AND PROSPECTS FOR FURTHER IMPROVEMENTS}

We show in Table \ref{tab:res} examples of the results for $\alpha_s(M_Z)$
based on the best determinations at various mass scales, both in the Standard
Model and in the presence of a light gluino and bottom squark.  As mentioned
earlier, for a gluino in the 12 to 16 GeV mass range, values of $\delta
\alpha_s(M_Z)$ due to this latter scenario range from $\simeq 0.015$ to $\simeq
0.013$ when extrapolating from $m_b$ to $M_Z$, so we shall quote their effect
as $0.014 \pm 0.001$.

\begin{table}
\caption{Values of $\alpha_s(M_Z)$ based on determinations at different
mass scales, in the Standard Model (1) and in the presence of a light gluino
and bottom squark (2).
\label{tab:res}}
\begin{center}
\begin{tabular}{c c c c} \hline \hline
Source  & $Q$ (GeV) & (1) & (2) \\ \hline
$\tau$  &    1.78   & $0.118~(0.121?) \pm (>)0.003$ & (a) \\
  DIS   &  $\sim 3$ & $0.1166 \pm (>)0.0022$ & $0.130 \pm (>)0.003$ \\
Lattice &  $\sim 5$ & $0.121 \pm (>)0.003$ & $0.135 \pm (>)0.003$ \\
$\Gamma_h(Z)$ & 91.2 & $0.123 \pm 0.005$ & $(0.123-0.131) \pm 0.005$ \\
Ev.\ shapes & $> M_Z$ & $0.123 \pm 0.006$ & Unknown \\ \hline \hline
\end{tabular}
\end{center}
\leftline{(a) See Sec.\ III A.  Extrapolation from such a low $Q$ is risky
in our opinion.}
\end{table}

\begin{figure}
\includegraphics[height=12cm]{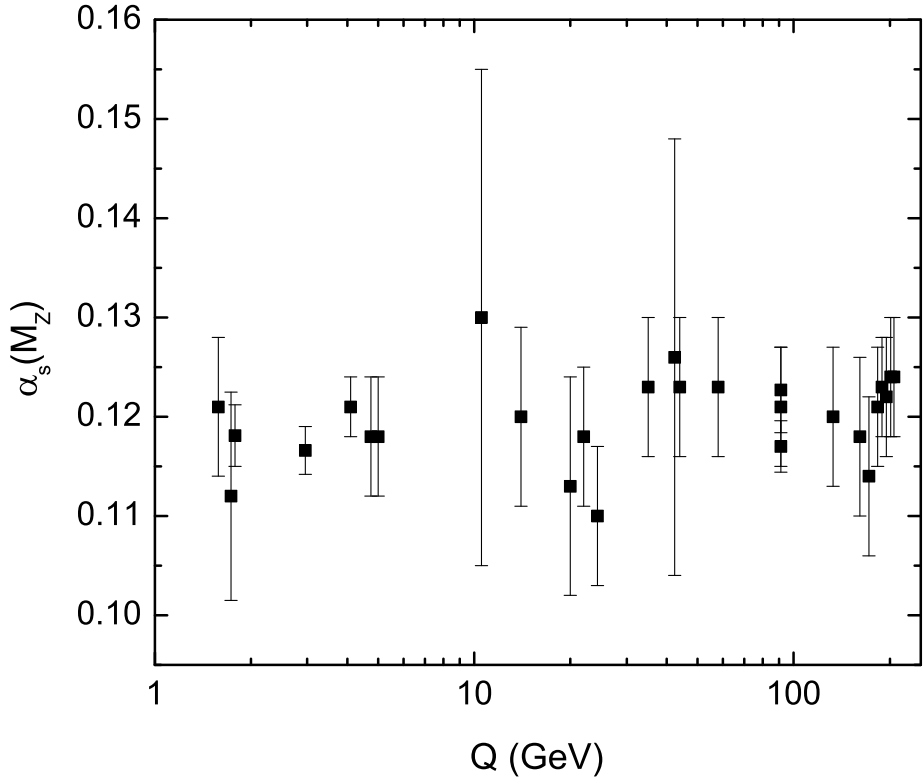}
\caption{Values of $\alpha_s(M_Z)$ as determined at various values of $Q$,
based on Standard Model evolution. From Ref.\ \cite{Bethke:2002rv}.
\label{fig:qsq}}
\end{figure}

Table \ref{tab:res} presents a rather unsatisfactory situation at present, in
our view.  No clear-cut decision is possible in favor of either the Standard
Model or the light gluino/bottom squark scenario.  In Fig.\ \ref{fig:qsq} we
show values of $\alpha_s(M_Z)$ extracted from determinations at various values
of $Q$ \cite{Bethke:2002rv}.  A straight line, corresponding to the Standard
Model, clearly provides an excellent fit, while we have shown that the
best-measured values of $\alpha_s$ are also compatible with the light
gluino/bottom squark hypothesis.

We expect that some of the indeterminacy should be reduced when results of
fully unquenched lattice calculations appear, reducing the error on the
extrapolated coupling constant to $\Delta \alpha_s(M_Z) = \pm 0.002$ or less.
However, further reduction of uncertainty will require improved determinations
either at the $Z$ mass (particularly of $\Gamma_{\rm tot}(Z)$ and $R_b$) or
above it (extrapolated down to $M_Z$).  For the latter case, a calculation is
needed for the effect of the light gluino/bottom squark proposal on hadronic
event shapes.

\section{SUMMARY \label{sec:sum}}

We have outlined the current status of the scale-dependence of the strong
fine-structure constant $\alpha_s$ and the light it can shed on the hypothesis
of a light gluino and bottom squark.  No conclusion is possible at present
regarding this hypothesis with CP violating phases vis-\`a-vis the Standard
Model.  Improvements that will permit a more clear-cut test include refinement
of lattice calculations, reduction of errors based in the hadronic and $b \bar
b$ widths of the $Z$, and possibly more precise determinations based on event
shapes in high-energy $e^+ e^-$ collisions.  Of course, direct searches for
light gluinos and bottom squarks will play a key role, but that is another
story.

{\it Note added:} After this work was finished, we received a paper
\cite{Cheung:2002} considering the $Z \to b {\tilde b^*} {\tilde g} + b^*
{\tilde b} {\tilde g}$ channel, whose partial width was estimated to be of
order $10^{-3}$ GeV in the gluino mass range of interest to us.  This positive
contribution may partially cancel with the negative SUSY contribution to
$\Gamma(Z \to b \bar b)$ in both CP-conserving and CP-violating cases
\cite{Baek:2002xf,Cao:2001rz} and, therefore, brings down our estimate of
$\alpha_s$ extracted at $M_Z$ in Table \ref{tab:res}, where maximal CP
violation is assumed.

\section*{ACKNOWLEDGMENTS}

We would like to thank S.~w.~Baek, E.~Berger, A.~Djouadi, M.~Drees,
A.~Leibovich, A.~Kagan, A.~ Kronfeld, S.~Martin, T.~Tait, and C. Wagner for
useful discussions.  This work was supported in part by the U.\ S.\ Department
of Energy, High Energy Physics Division, through Grant No.\ DE-FG02-90ER-40560
and under Contract W-31-109-ENG-38.

\end{document}